\documentclass[12pt]{article}

\newcommand{\commentstarts}{\begin{centering}
\hspace{-1pt}\vrule\vrule
\begin{minipage}[t]{0.03\linewidth}
\hspace{0.025\linewidth}
\end{minipage}
\begin{minipage}[t]{0.95\linewidth}}
\newcommand{\commentends}{\end{minipage}
\end{centering}
\vspace{7pt}
}

\usepackage{graphicx}
\usepackage{alltt}
\usepackage{amsmath}
\usepackage{amssymb}
\usepackage{hyperref}

\begin{document} 

\newcounter{Theorems}
\setcounter{Theorems}{0}

\newcounter{Definitions}
\setcounter{Definitions}{0}

\begin{titlepage}
\begin{flushright}

\end{flushright}

\begin{center}
{\Large\bf $ $ \\ $ $ \\
A geometrical point of view on linearized beta-deformations
}\\
\bigskip\bigskip\bigskip
{\large Andrei Mikhailov${}^{\dag}$ and Segundo P. Mili\'{a}n}
\\
\bigskip\bigskip
{\it Instituto de F\'{i}sica Te\'orica, Universidade Estadual Paulista\\
R. Dr. Bento Teobaldo Ferraz 271, 
Bloco II -- Barra Funda\\
CEP:01140-070 -- S\~{a}o Paulo, Brasil\\
}

\vskip 1cm
\end{center}

\begin{abstract}
It is known that the supermultiplet of beta-deformations of ${\cal N}=4$ supersymmetric Yang-Mills
theory can be described in terms of the exterior product of two adjoint representations
of the superconformal algebra. We present a super-geometrical interpretation of this fact, by
evaluating the deforming operator on some special coherent states in the space of supersingletons.
We also discuss generalization of this approach to other finite-dimensional deformations of the
${\cal N}=4$ supersymmetric Yang-Mills theory.
\end{abstract}

\vfill
{\renewcommand{\arraystretch}{0.8}%
\begin{tabular}{rl}
${}^\dag\!\!\!\!$ 
& 
\footnotesize{on leave from Institute for Theoretical and 
Experimental Physics,}
\\    
&
\footnotesize{ul. Bol. Cheremushkinskaya, 25, 
Moscow 117259, Russia}
\\
\end{tabular}
}

\end{titlepage}

\tableofcontents 

\section{Introduction}\label{sec:Introduction}
AdS/CFT correspondence has been under vigorous study for more than 20 years, bringing
important results. Historically the first application was to compare the BPS (or ``protected'')
states of the $N=4$ supersymmetric Yang-Mills theory to the solutions of Type IIB supergravity (SUGRA)
linearized near the ``background solution'' $AdS_5\times S^5$. Both are representations of Lie
super-algebra ${\bf psu}(2,2|4)$. The fact that they precisely match has been firmly established, 
see \cite{Aharony:1999ti} and references therein. However, the structure of these representations has not been
sufficiently well studied. We understand well those representations which are unitary. Those are
irreducible, generated by highest weight vectors (UIR) \cite{Gunaydin:1998sw}.
But not all interesting representations
are irreducible (and even for UIRs, it would be good to have several alternative descriptions).
For example, there are finite-dimensional representations and, to the best of our
knowledge, it is not clear if they are irreducible or not\footnote{finite dimensional
representations of Lie superalgebras are not necessarily semisimple}. Understanding, from various points
of view, the structure of the representations appearing in AdS/CFT would be useful. In particular, it is needed
for constructing the string worldsheet massless vertex operators, along the lines of \cite{Mikhailov:2011si,Mikhailov:2011af,Berkovits:2012ps,Mikhailov:2013dp}. (The observation of \cite{Mikhailov:2011si} was that, to construct the
vertex, it is enough to know the structure of the representation of ${\bf psu}(2,2|4)$ in which the
corresponding state transforms.) And an efficient construction of massless vertex operators would,
in turn, allow the computation of the SUGRA scattering amplitudes on $AdS_5\times S^5$ in the pure spinor formalism \cite{Berkovits:2000fe}.

In this paper we will suggest a point of view on the {\em finite-dimensional} represenations.
The idea is to probe them by evaluating the corresponding deformations of the free Yang-Mills
action on certain solutions of free field equations. The result is some generalized function
of the parameters of those free solutions. This generalized function is manifestly
supersymmetric and has a nice super-geometrical interpretation. We discuss mostly the case of the so-called beta-deformation \cite{Leigh:1995ep}. The structure of the corresponding
representation is known from the supergravity analysis\footnote{Beta-deformations were discovered
  in \cite{Leigh:1995ep}, but their structure as representations of ${\bf psu}(2,2|4)$ was not
  studied. It was computed using AdS/CFT in \cite{Bedoya:2010qz}.}\cite{Bedoya:2010qz}. Here we
make this structure transparent on the field theory side. Then we apply our method to other
finite-dimensional representations. Under the assumption (unproven) that they are irreducible,
we identify them with particular Young diagramms. 

\vspace{10pt}
\noindent We will now proceed to describing our results. 

\paragraph     {Deformations transforming in finite-dimensional representations}

Consider single trace deformations of ${\cal N}=4$ supersymmetric Yang-Mills theory (SYM):
\begin{equation}\label{DeformationsOfSYM}
S_{\rm SYM} \;\mapsto \; S_{\rm SYM} + \varepsilon \int d^4x U
\end{equation}
where $U$ is a single-trace operator. 
According to AdS/CFT \cite{Aharony:1999ti}, they correspond to some deformations of the Type IIB superstring theory
on $AdS_5\times S^5$. Let us restrict ourselves to infinitesimal deformations, {\it i.e.} compute
only to the first order in the deformation parameter $\varepsilon$. Besides $\varepsilon$, there are two
other parameters: the Yang-Mills coupling constant $g_{YM}$ and the number of colors $N$. In this 
paper, we will consider $N$ being very large. As for the $g_{YM}$, there are two opposite limits: 
the limit $g_{YM}=0$ of free ${\cal N}=4$ super Yang-Mills theory (SYM) and the strong coupling 
limit. In the weak coupling limit we can do perturbative calculations in SYM, and in the strong 
coupling limit we can use the superstring theory on the classical supergravity (SUGRA) background 
$AdS_5\times S^5$.

Both SYM theory and superstring on $AdS_5\times S^5$ are invariant under the superconformal group 
$PSU(2,2|4)$, and it is natural to ask how the deformations of the form (\ref{DeformationsOfSYM}) transform 
under this group. In particular, some deformations of the form (\ref{DeformationsOfSYM})
transform in finite-dimensional representations \cite{Mikhailov:2011af}. 
As an example of a finite-dimensional representation, consider $U$ of Eq. (\ref{DeformationsOfSYM}) of the form \cite{Mikhailov:2011af}:
\begin{equation}\label{ChiralPrimaryDeformation}
\int d^4x \; \mbox{tr}(\Phi_1+i\Phi_2)^{n+4} \;+\; \mbox{c.c.}
\end{equation}
where $\Phi_1,\ldots,\Phi_6$ are the scalar fields of the ${\cal N}=4$-super-Yang-Mills theory. Consider
the linear space of all deformations obtained from Eq. (\ref{ChiralPrimaryDeformation}) by acting with all possible
polynomials of generators of $psu(2,2|4)$. In the free theory, this gives a finite-dimensional representation.
Let us call it $E^{\rm free}_{n+4}$.\marginpar{$E_{n+4}$}

Most of finite-dimensional deformations of the free theory combine into infinite-dimensional
representations in the quantum interacting theory. But $E^{\rm free}_{n+4}$ stays finite-dimensional.
We will call it just $E_{n+4}$.
This follows from the finiteness of the operator $\mbox{tr}(\Phi_1+i\Phi_2)^{n+4}$ and its descendants.
Finiteness implies that the conformal transformations
of this operators are same in quantum theory as in classical theory (no countertems $\Rightarrow$
no anomalous dimension).
Therefore, the subspace generated by acting on Eq. (\ref{ChiralPrimaryDeformation}) with the conformal generators is the
same in classical and quantum theory. In particular, it is finite-dimensional. To generate
the full representation, it remains to act with supersymmetry and superconformal generators.
However, those are nilpotent modulo elements of $su(2,2)\oplus su(4)$, and therefore cannot change the
property of the representation being finite-dimensional.

We will now argue that  $E_{n+4}$ is isomorphic to the representation $E_{n+4}^{\rm free}$ which is generated
by acting with $psu(2,2|4)$ on  Eq. (\ref{ChiralPrimaryDeformation}) in the {\em free} ${\cal N}=4$ super-Yang-Mills theory.
Indeed, let us consider the {\em chiral primary representation}, which is defined as the
infinite-dimensional representation generated by the local operator $\mbox{tr}(\Phi_1 + i\Phi_2)^{n+4}(0)$
by acting on it with $psu(2,2|4)$. This is a unitary representation. Let us call it $F_{n+4}$.
\marginpar{$F_{n+4}$}
Similarly, there is a free version $F_{n+4}^{\rm free}$ which is defined in the same way as
$F_{n+4}$, but in the free theory. We start by pointing out that $F_{n+4}$ is isomorphic
to $F^{\rm free}_{n+4}$, because both are highest weight unitary irreducible representations of $psu(2,2|4)$
with the same highest weight  $\mbox{tr}(\Phi_1 + i\Phi_2)^{n+4}(0)$. On the other hand, our finite-dimensional
representations $E_{n+4}$ and $E^{\rm free}_{n+4}$ are fully determined
by $F_{n+4}$ and $F^{\rm free}_{n+4}$, respectively, in the following way.
Elements of $E_{n+4}$ are of the form $\int d^4x\;\rho(x)\;{\cal O}(x)$
where ${\cal O}$ is an element of $F_{n+4}$ (a local operator), but inserted at the point $x$
instead of $0$. When we act on  $\int d^4x\;\rho(x)\;{\cal O}(x)$ with an element of $psu(2,2|4)$,
the result is determined by how this element acts on ${\cal O}(x)$,
{\it i.e.} by the structure of $F_{n+4}$. The isomorphism $F_{n+4}\stackrel{\simeq}{\longrightarrow}F^{\rm free}_{n+4}$ implies a map
${\cal O}\mapsto {\cal O}^{\rm free}$ commuting with the action of $psu(2,2|4)$. This establishes an isomoprhism
$E_{n+4}\stackrel{\simeq}{\longrightarrow}E^{\rm free}_{n+4}$.

Still, it is not clear to us if $E_{n+4}$ is irreducible or not. If it is irreducible, then
considerations of \cite{Mikhailov:2011af} and Section \ref{sec:Generalization} of this paper suggest that it should correspond to the supersymmetric Young diagramms (see Section \ref{sec:Generalization} and   \cite{Mikhailov:2011af} and references there):
\begin{center}
\begin{minipage}[c]{0.94\linewidth}
\begin{center}
   \includegraphics[width=3in]{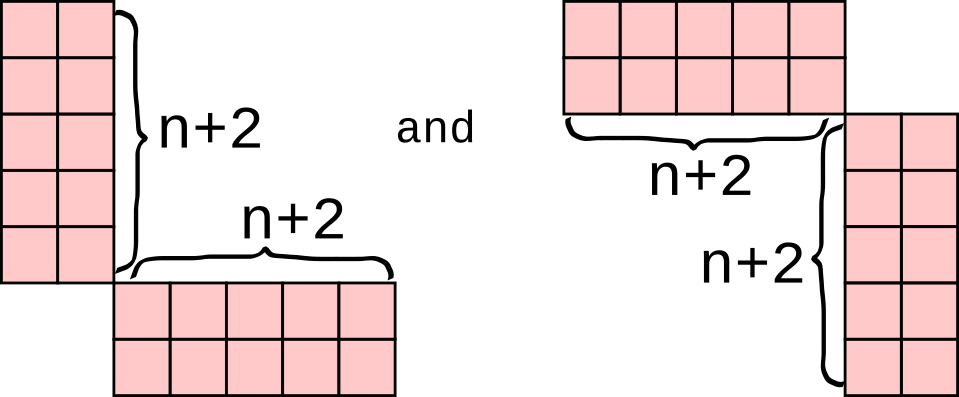}
\end{center}
\end{minipage}
\begin{minipage}[t]{0.05\linewidth}
    {\refstepcounter{equation}\label{two-diagramms}} 
    (\arabic{equation})
 \end{minipage}
\end{center}

\paragraph     {Beta-deformation}
In this letter we will mostly consider a particular example of a finite-dimensional representation:
the linearized $\beta$-deformation \cite{Leigh:1995ep,Bedoya:2010qz}. In some sense it corresponds to $n=-1$, but
the construction is different. If we literally take $n=-1$ in Eq. (\ref{ChiralPrimaryDeformation}), and start acting
by the generators of $psu(2,2|4)$, we obtain an infinite-dimensional representation.
However, there is a finite dimensional subrepresentation, which is generated by:
\begin{equation}
   \int d^4x\;\epsilon^{\alpha\beta}
   Q^{[-{1\over 2}, {1\over 2}, {1\over 2}]}_{\alpha}
   Q^{[-{1\over 2}, {1\over 2}, {1\over 2}]}_{\beta}\mbox{tr}(\Phi_1 + i\Phi_2)^3 \;+\; \mbox{c.c.}
\end{equation}
where $Q^{[-{1\over 2}, {1\over 2}, {1\over 2}]}_{\alpha}$ are supersymmetry generators with weight $\left[-{1\over 2}, {1\over 2}, {1\over 2}\right]$ under the elements
$\Phi_1{\partial\over\partial \Phi_2} - \Phi_2{\partial\over\partial \Phi_1}$, $\Phi_3{\partial\over\partial \Phi_4} - \Phi_4{\partial\over\partial \Phi_3}$ and $\Phi_5{\partial\over\partial \Phi_6} - \Phi_6{\partial\over\partial \Phi_5}$ of the  $su(4)$ Cartan
subalgebra. This deformation is preserved by  $Q^{[-{1\over 2}, {1\over 2}, {1\over 2}]}_{\alpha}$ and $Q^{[{1\over 2}, -{1\over 2}, -{1\over 2}]}_{\dot{\alpha}}$.
It starts with the cubic term, but also has quartic terms.
In the language of ${\cal N}=1$ superspace, the
cubic terms come from $\delta W'' \psi\psi$ and quartic terms come from $|(W + \delta W)'|^2$ terms,
where $W$ is the superpotential of the ${\cal N}=4$ Yang-Mills
in the ${\cal N}=1$ notations, and $\delta W$ its Leigh-Strassler deformation. Notice that the quartic
terms are proportional to $g_{\rm YM}$. In the free field limit the quartic terms are absent,
but the cubic terms survive and define a deformation of the {\em free} Yang-Mills theory.

It was found on the AdS side that the supermultiplet
of $\beta$-deformations is related to the wedge product of two copies of the adjoint representation
of ${\bf g} = psu(2,2|4)$ \cite{Bedoya:2010qz}:
\begin{equation}\label{Representation}
{({\bf g}\wedge {\bf g})_0\over {\bf g}}
\end{equation}
This {\em is} an irreducible representation. The subscript $0$ imposes the constraint of
zero internal commutator; the origin of this constraing on the AdS side was explained in \cite{Mikhailov:2012id,Mikhailov:2014qka,Wulff:2016tju}.
The geometrical origin of ${\bf g}\wedge {\bf g}$ is immediately visible on the AdS side; the 
corresponding vertex operator is essentially the wedge product of two global symmetry currents on the
string worldsheet.
But it is not immediately obvious why such structure would turn up on the field theory side.
In the free field limit the deforming operator $U$ is some quibic expression in elementary
fields. It was shown in \cite{Milian:2016xuy} that these expressions transform in the representation given by
Eq. (\ref{Representation}). In this letter we will present a geometrical construction making this result
more transparent. We will use twistor methods. First in Section \ref{sec:HarmonicSuperspace} we will review
the formalism of Harmonic
superspace, and how it describes classical solutions of free theory.
We then introduce in Section \ref{sec:CoherentWaveFunctions} some special coherent states, which are $\delta$-functions
with support on $\alpha$-planes. (The ideas of Sections \ref{sec:HarmonicSuperspace} and \ref{sec:CoherentWaveFunctions} are well known to experts,
but we could not find a reference with explicit formulas suitable for our needs.) In Section \ref{sec:betadef} 
we consider the evaluation of the beta-deformation term $\int d^4x \;U$ in Eq. (\ref{DeformationsOfSYM}) on the formal 
sum of coherent states, also known as ``perturbiner''. 
For our approach it is essential to work in space-time signature $2+2$. In this case
the evaluation gives a well-defined generalized function of the parameters of our coherent states.
We explicitly describe this generalized function and study its properties.
In Section \ref{sec:Generalization} we conjecture a similar description for the deformations of the
form Eq. (\ref{ChiralPrimaryDeformation}) with $n\geq 0$.

\vspace{10pt}
\noindent
It should be possible to drive this results using twistor string theory \cite{Witten:2003nn,Berkovits:2004hg}, from a
generalization of the results of\footnote{We would like to thank the Referee for pointing
  out to us these references} \cite{Kulaxizi:2004pa,Kulaxizi:2005qt,Gao:2006mw}. Notice that twistor string theory requires complex twistor suprspace ${\bf CP}^{3|4}$, while
for our considerations here real twistors ${\bf RP}^{3|4}$ are enough ({\it cp.} Section 3.2 of \cite{Witten:2003nn}).

\vspace{10pt}
\noindent
Our approach can be interpreted as classifying the possible deformations of the action by
looking at their effect on the scattering amplitudes. In fact, Eqs. (\ref{EvaluationEqualsDelta}) and (\ref{Generalization}) can be
interpreted as the deformation of the $k$-point scattering amplitude at the minimal value of $k$
for which the deformation is nonzero.
Conceptually similar approach was used in \cite{Elvang:2010jv} for classification of counterterms in supergravity.

\section{Harmonic superspace}\label{sec:HarmonicSuperspace}

\subsection{Twistors and Grassmannian}\label{sec:TwistorsAndGrassmannian}
We will work with the space-time signature $2+2$. With this signature the superconformal group is $PSL(4|4,{\bf R})$.
Following\footnote{although with our choice of signature we cannot directly use complex geometry;
we will replace complex analyticity with some polynomiality condition.} \cite{MR1632008} we will think about the four-dimensional ${\cal N}=4$ superspace as a Grassmannian
manifold $M = {\rm Gr}(2|2,{\bf R}^{4|4})$, which parametrizes subspaces ${\bf R}^{2|2}\subset {\bf R}^{4|4}$.

We will consider the twistor space ${\bf T}={\bf R}^{4|4}$; elements of $\bf T$ are vectors parametrized by 
even $\lambda_{\dot{\alpha}},\mu^{\alpha}$ and odd $\zeta^I$:\marginpar{$\bf T$}\marginpar{$\lambda,\,\mu$}
\begin{align}\label{CoordinatesInTwistorSpace}
Z\;=\;& (\mu^1,\mu^2,\lambda_{\dot{1}},\lambda_{\dot{2}},\zeta_1,\zeta_2,\zeta_3,\zeta_4)^T
\end{align}
Here $(\ldots)^T$ means transposition; we think of $Z$ as a ``column-vector'', but write it as a
row in Eq. (\ref{CoordinatesInTwistorSpace}) for typographic reasons.
Points of $M$ are $2|2$-planes $L\subset {\bf T}$. They can be parametrized by coordinates $x,\theta,\tilde{\theta},u$:
\begin{align}\label{BasisOfL}
\mbox{\tt\small basis of } L^{\perp}\subset {\bf T}':\quad
\begin{array}{rccccccccl}
( & 1 & 0 & x^{1\dot{1}} & x^{1\dot{2}} & \;\;
\theta^{11} & \theta^{12} & \theta^{13} & \theta^{14} & ) 
\cr
( & 0 & 1 & x^{2\dot{1}} & x^{2\dot{2}} & \;\;
\theta^{21} & \theta^{22} & \theta^{23} & \theta^{24} & ) \cr
( & \tilde{\theta}_{11} & \tilde{\theta}_{12} & \tilde{\theta}_1^{\dot{1}} & 
\tilde{\theta}_1^{\dot{2}} &
u_1^1 & u_1^2 & u_1^3 & u_1^4 &) \cr
( &\tilde{\theta}_{21} & \tilde{\theta}_{22} & \tilde{\theta}_2^{\dot{1}} & 
\tilde{\theta}_2^{\dot{2}} & 
u_2^1 & u_2^2 & u_2^3 & u_2^4 &)
\end{array}
\end{align}
In these coordinates the condition $Z\in L$ is that each row-vector from the set of four
vectors (\ref{BasisOfL}) has zero scalar product with the column-vector $Z$. 
These coordinates are redundant. The redundancy can be described as some vector fields
whose flows do not change $L$, for example: 
\begin{align}
  & A^{i\alpha}\left(
    {\partial\over\partial \tilde{\theta}_{i\alpha}} + 
    x^{\alpha\dot{1}}{\partial\over\partial\tilde{\theta}^{\dot{1}}_{i}} +
    x^{\alpha\dot{2}}{\partial\over\partial\tilde{\theta}^{\dot{2}}_{i}} +
    \theta^{\alpha I} {\partial\over\partial u^I_{i}}
    \right)
  \\    
  & A^K_L\left(
    \tilde{\theta}_{KA}{\partial\over\partial\tilde{\theta}_{LA}} +
    \tilde{\theta}^{\dot A}_K{\partial\over\partial\tilde{\theta}^{\dot{A}}_L} +
    u^I_K{\partial\over\partial u^I_L}
    \right)
    \label{MixingThirdAndFourth}
\end{align}
for any constant $A^{i\alpha}$ and $A^K_L$.
This can be used to put $\tilde{\theta}_{i\alpha} = 0$:
\begin{equation}
\tilde{\theta}_{11} = \tilde{\theta}_{12} = \tilde{\theta}_{21} = \tilde{\theta}_{22} = 0
\end{equation}
To summarize:
\begin{align}
\mbox{\tt\small twistors}  =\;& {\bf T} = {\bf R}^{4|4}
\\    
\mbox{\tt\small space-time}=\;& M = \mbox{Gr}(2|2,{\bf T})
\end{align}
A special role is played by the $2|2$-dimensional tautological vector bundle over $M$ which will
be denoted ${\cal S}_M$. The fiber of ${\cal S}_M$ over $L\subset M$ is $L$ itself. \marginpar{${\cal S}_M$} 

Also, $\bf PT$ will denote the projective super-twistor space, {\it i.e.}
$Z$ modulo rescaling: $Z\simeq \kappa Z$. \marginpar{$\bf PT$}

\subsection{Berezinian and some notations}\label{sec:Superdeterminant}
For any vector bundle $\cal V$, we define two line bundles, which we call  $\mbox{Ber}\,\cal V$ and $|\mbox{Ber}|\,\cal V$.
\marginpar{$\mbox{Ber}$} \marginpar{$|\mbox{Ber}|$}
Sections of $\mbox{Ber}\,\cal V$ are functions of the bases of the fiber $V$, satisfying the property:
\begin{equation}
\sigma(\{m^J_Ie_J\}) = \mbox{SDet}\,m\;\;\sigma(\{e_I\})
\end{equation}
where for a supermatrix $m = \left(\begin{array}{cc} A & B \cr C & D \end{array}\right)$:
\begin{equation}
\mbox{SDet}\;m \;=\;\mbox{det}(A - BD^{-1}C)\;(\mbox{det}\;D)^{-1}
\end{equation}
Similarly, section of $|\mbox{Ber}|\,V$ satisfy:
\begin{equation}
\sigma(\{m^J_Ie_J\}) =\; \mbox{sign}(\mbox{det}\,A)\; \mbox{SDet}\,m\;\; \sigma(\{e_I\})
\end{equation}
The characteristic property of  $|\mbox{Ber}|\,\cal V$ is that a section of  $|\mbox{Ber}|\,\cal V$
defines an operation of {\em integration along the fiber}:
\begin{equation}
   \left[\begin{array}{c}
           \mbox{\tt\small functions on $\cal V$ }
           \cr
           \mbox{\tt\small with compact support}
           \cr
           \mbox{\tt\small on each fiber $V$}
         \end{array}
      \right]
      \stackrel{\int}{\longrightarrow}
      \left[\begin{array}{c}
              \mbox{\tt\small functions} 
              \cr
              \mbox{\tt\small on the base}
            \end{array}\right]
\end{equation}
It seems logical to call sections of  $\mbox{Ber}\,\cal V$ ``integral forms on the fiber'', and section
of  $|\mbox{Ber}|\,\cal V$  ``volume elements on the fiber''.

We will denote $(\mbox{Ber}\,{\cal V})^{-1}$ the line bundle dual to $\mbox{Ber}\,{\cal V}$,
{\it i.e.} same as $(\mbox{Ber}\,{\cal V})'$,\marginpar{$\mbox{Ber}^{-1}$}
and  $|\mbox{Ber}\,{\cal V}|^{-1} = |\mbox{Ber}\,{\cal V}|'$.

\subsection{Action of $GL(4|4,{\bf R})$ transformations on twistors}
The twistor vector space ${\bf T}$ should be considered a fundamental representation
of  $GL(4|4,{\bf R})$. We have:
\begin{equation}
GL(4|4)/{\rm center} \;=\; PGL(4|4) \;=\; PSL(4|4,{\bf R}) \rtimes U(1)_R
\end{equation}
The $u(1)_R$ acts as follows:
\begin{equation}
R Z = \left(\begin{array}{rr} 
      -{\bf 1}_{4\times 4} & {\bf 0}_{4\times 4} \cr 
      {\bf 0}_{4\times 4} & {\bf 1}_{4\times 4}
\end{array}
\right)Z
\end{equation}
In terms of the parametrization (\ref{CoordinatesInTwistorSpace}): 
\begin{equation}
R = - \lambda_{\dot{\alpha}}{\partial\over\partial\lambda_{\dot{\alpha}}} 
- \mu^{\alpha}{\partial\over\partial\mu^{\alpha}} 
+ \zeta^I{\partial\over\partial\zeta^I}
\end{equation}

\subsection{Solutions of free classical super-Maxwell equations from twistor space}
In the free field limit a classical solution of the super-Yang-Mills theory with gauge group
$U(N)$  can be obtained as an $N\times N$ matrix whose entries are classical solutions
of the super-Maxwell equations. For our purpose it is enough to consider free solutions
of the form:
\begin{equation}
   \left[
      \begin{array}{c}
        \mbox{\tt\small solution of }
        \cr
        \mbox{\tt\small super-Maxwell equations}
      \end{array}
   \right]
   \times
   \left[
      \begin{array}{c}
        N\times N
        \cr
        \mbox{\tt\small matrix}
      \end{array}
   \right]
   \label{MaxwellTimesMatrix}
\end{equation}
In this Section we will explain the correspondence between
solutions of free classical super-Maxwell equations on $M$ and sections of $\left(|\mbox{Ber}|\,{\cal S}_M\right)^{-1}$
which are polynomial in the coordinates $u^I_{i}$ defined in Eq. (\ref{BasisOfL}).  
The choice of coordinates $u_{i}^I$ is somewhat arbitrary, but the statement of polynomial dependence 
does not depend on it; in particular this requirement 
is $PSL(4|4,{\bf R}) \rtimes U(1)_R$-invariant.

\vspace{10pt}
\noindent
Notice that $|\mbox{Ber}|{\cal S}_M = \left(|\mbox{Ber}|({\bf T}/{\cal S}_M)\right)^{-1}
= |\mbox{Ber}|({\bf T}/{\cal S}_M)' = |\mbox{Ber}|{\cal S}^{\perp}_M$.
Those four vectors listed in Eq. (\ref{BasisOfL}) are the basis of ${\cal S}_M^{\perp}$. 

\vspace{10pt}
\noindent
Suppose that we are given a section (here $\Gamma(\mbox{\tt line bundle})$ denotes the space
of sections):
\begin{equation}\label{TypeOfSigma}
\sigma\in \Gamma\left(\left(|\mbox{Ber}|\,{\cal S}_M\right)^{-1}\right)
\end{equation}
which is a polynomial in $u$.
The condition that $\sigma$ is annihilated by the vector fields defined in Eq. (\ref{MixingThirdAndFourth}) with $A^K_L$
satisfying $A^K_K=0$ implies that
it can depend on $\tilde{\theta}$ and $u$ only through the expressions $\tilde{\theta}^{\dot{\alpha}}_{[1}u^I_{2]}$, $u^I_{[1}u^J_{2]}$ and $\tilde{\theta}^{\dot{\alpha}}_{[1}\tilde{\theta}^{\dot{\beta}}_{2]}$. Since
we require $\sigma$ to be a polynomial in $u$, it must be then a polynomial of these three
expressions. Moreover, $\sigma$ should have charge $2$ under rescaling of $(\tilde{\theta},u)$, therefore this
polynomial is actually a linear function:
\begin{align}
\label{def-of-sigma}
\sigma = \tilde{\theta}^{\dot{\alpha}}_{[1}\tilde{\theta}^{\dot{\beta}}_{2]}{\bf F}^+_{\dot{\alpha}\dot{\beta}}(x,\theta)
+ \tilde{\theta}^{\dot{\alpha}}_{[1}u^I_{2]}{\bf \Psi}_{\dot{\alpha}I}(x,\theta)
+ u^I_{[1}u^J_{2]} {\bf \Phi}_{IJ}(x,\theta)
\end{align}
Consider the condition that $\sigma$ does not change when we add to first and second vectors
of the basis of Eq. (\ref{BasisOfL}) a linear combination of the third and fourth vectors.
It implies that $\sigma$ should be annihilated by the following vector fields $V_{i\alpha}$:
\begin{align}
V_{i\alpha}\;\sigma\;=\;& 0
\label{VSigmaIsZero}\\      
\mbox{\tt\small where } V_{i\alpha} \;=\;& \tilde{\theta}_i^{\dot{\beta}} {\partial\over\partial x^{\alpha\dot{\beta}}} +
u_i^I{\partial\over\partial\theta^{\alpha I}}
\end{align}
In the rest of this Section, we will first prove that Eq. (\ref{VSigmaIsZero}) implies 
the following equations:
\begin{align}
\label{const1-in-F+}
\epsilon^{\dot{\gamma}\dot{\beta}}
{\partial\over\partial x^{\alpha\dot{\gamma}}}{\bf F}^+_{\dot{\alpha}\dot{\beta}}(x,\theta)\;=\;& 0,
\\   
\label{const2-in-F+}
{\partial\over\partial\theta^{\alpha I}}{\bf F}^+_{\dot{\alpha}\dot{\beta}}(x,\theta)\;=\;& 
{\partial\over\partial x^{\alpha(\dot{\beta}}}{\bf \Psi}_{\dot{\alpha})I}(x,\theta),
\\  
\label{eqn-for-Psi}
\epsilon^{\dot{\alpha}\dot{\beta}}
{\partial\over\partial x^{\alpha\dot{\alpha}}}{\bf \Psi}_{\dot{\beta}I}(x,\theta)\;=\;& 0,
\\  
\label{const1-for-Psi}
{\partial\over\partial\theta^{\alpha(I}}{\bf \Psi}_{J)\dot{\alpha}}(x,\theta)\;=\;& 0,
\\    
\label{const2-for-Psi}
{\partial\over\partial\theta^{\alpha[I}}{\bf \Psi}_{J]\dot{\alpha}}(x,\theta)\;=\;& 
{\partial\over\partial x^{\alpha\dot{\alpha}}}{\bf \Phi}_{IJ}(x,\theta), 
\\
\label{const-for-Phi}
{\partial\over\partial \theta^{\alpha P}}{\bf \Phi}_{IJ}(x,\theta)\;=\;&{\partial\over\partial\theta^{\alpha[P}}{\bf \Phi}_{IJ]}(x,\theta),
\end{align}
 and then show that 
Eqs.~\eqref{const1-in-F+}-~\eqref{const-for-Phi} imply that the $\theta$-expansion 
of ${\bf F}^{+}_{\dot\alpha\dot\beta}(x,\theta),\; {\bf \Psi}_{J\dot\alpha}(x,\theta)$ and ${\bf \Phi}_{IJ}(x,\theta)$ can be expressed through solutions of free 
super-Maxwell theory. 

\paragraph     {Proof that Eq. (\ref{VSigmaIsZero}) implies Eqns.~\eqref{const1-in-F+}-~\eqref{const-for-Phi}}
Consider the expansion of the left hand side of Eq. (\ref{VSigmaIsZero}) in powers of $\tilde{\theta}$.
\begin{itemize}
 \item At the zeroth order in $\tilde\theta$, the term that contributes is
 \begin{align}
  u^P_mu^I_{[1}u^J_{2]}{\partial\over\partial\theta^{\alpha P}}{\bf \Phi}_{IJ}(x,\theta).
 \end{align}
 Its vanishing implies Eq. (\ref{const-for-Phi})
\item At the linear order in $\tilde\theta$, we have 
\begin{align}
\label{1tildetheta}
 - u^I_m\tilde\theta^{\dot\alpha}_{[1}u^J_{2]}{\partial\over\partial\theta^{\alpha I}}{\bf \Psi}_{\dot\alpha J}(x, \theta)+\tilde\theta^{\dot\beta}_mu^I_{[1}u^J_{2]}\partial_{\alpha\dot\beta}{\bf \Phi}_{IJ}(x,\theta),
\end{align}
where $\partial_{\alpha\dot\beta}$ denotes ${\partial\over\partial x^{\alpha\dot\beta}}$. Notice:
\begin{align}
 u^{I}_iu^J_j\;=\; \epsilon_{ij}u_{[1}^{[I}u_{2]}^{J]}+u^{(I}_{(i}u^{J)}_{j)},
\end{align}
Therefore the terms of Eq.~\eqref{1tildetheta} can be written as:
\begin{align}
 u^I_{[1}u^J_{2]}\tilde\theta^{\dot\alpha}_m\left( - {\partial\over\partial\theta^{\alpha [I}}{\bf \Psi}_{J]\dot\alpha}(x,\theta)
 + \partial_{\alpha\dot\alpha}{\bf \Phi}_{IJ}(x,\theta)\right) - u^I_{(1}u^J_{2)}\tilde\theta^{\dot\alpha}_{m}{\partial\over\partial\theta^{\alpha(I}}{\bf \Psi}_{J)\dot\alpha}(x,\theta). 
\end{align}
The vanishing of this expression implies Eqs. (\ref{const2-for-Psi}) and (\ref{const1-for-Psi}).
\item The terms of the second order in $\tilde\theta$ are
\begin{align}
\label{2tildetheta}
 u^I_m\tilde\theta^{\dot\alpha}_{[1}\tilde\theta^{\dot\beta}_{2]}{\partial\over\partial\theta^{\alpha I}}{\bf F}^{+}_{\dot\alpha\dot\beta}(x,\theta)
 +\tilde\theta^{\dot\beta}_m\tilde\theta^{\dot\alpha}_{[1}u^I_{2]}\partial_{\alpha\dot\beta}{\bf \Psi}_{\dot\alpha I}(x,\theta),
 \end{align}
using
\begin{align}
 \tilde\theta^{\dot\beta}_m\tilde\theta^{\dot\alpha}_k\;=\; 
\epsilon^{\dot\beta\dot\alpha}\tilde\theta^{[1}_{(m}\tilde\theta^{2]}_{k)}
 +\epsilon_{mk}\tilde\theta_{[1}^{(\dot\beta}\tilde\theta_{2]}^{\dot\alpha)},
\end{align}
Eqn.~\eqref{2tildetheta} can be written as
\begin{align}
u^I_m\tilde\theta^{\dot\alpha}_{[1}\tilde\theta^{\dot\beta}_{2]}\left({\partial\over\partial\theta^{\alpha I}}{\bf F}^{+}_{\dot\alpha\dot\beta}(x,\theta)
-\partial_{\alpha(\dot\beta}{\bf \Psi}_{\dot\alpha)I}\right) + \tilde\theta^{[1}_{(1}\tilde\theta^{2]}_{2)}u^I_m\left(\epsilon^{\dot\beta\dot\alpha}\partial_{\alpha\dot\beta}{\bf \Psi}_{\dot\alpha I}(x,\theta)\right),  
\end{align}
This is only zero when both terms vanish separately, implying Eqs. (\ref{const2-in-F+}) and (\ref{eqn-for-Psi}).
\item At third order in $\tilde\theta$, the term that contributes is:
\begin{align}
 T_{\dot\gamma\dot\beta\dot\alpha}\tilde\theta^{\dot\gamma}_m\tilde\theta^{\dot\beta}_{[1}\tilde\theta^{\dot\alpha}_{2]},
\end{align}
where $T_{\dot\gamma\dot\beta\dot\alpha}\;=\;\partial_{\alpha\dot\gamma}{\bf F}^+_{\dot\alpha\dot\beta}(x,\theta)$. The vanishing of this expression implies that  $T_{\dot\gamma\dot\beta\dot\alpha}$ must 
be  totally symmetric, and therefore Eq. (\ref{const1-in-F+}).
\end{itemize}
This concludes the derivation of the constraints \eqref{const1-in-F+}-~\eqref{const-for-Phi}
on the superfields ${\bf F}^{+}_{\dot\alpha\dot\beta}(x,\theta),\; {\bf \Psi}_{J\dot\alpha}(x,\theta)$ and  ${\bf \Phi}_{IJ}(x,\theta)$. 

\paragraph     {Solution to Eqns.~\eqref{const1-in-F+}-~\eqref{const-for-Phi}}
Now we will prove that the components of the $\theta$-expansion of ${\bf F}^{+}_{\dot\alpha\dot\beta}(x,\theta),\; {\bf \Psi}_{J\dot\alpha}(x,\theta)$ and  
${\bf \Phi}_{IJ}(x,\theta)$ can be expressed through solutions of free super-Maxwell theory. 

\vspace{10pt}
\noindent
Eq.~\eqref{const-for-Phi} implies:
\begin{align}
 \partial_{\gamma K}\partial_{\beta L}\partial_{M\alpha}{\bf \Phi}_{IJ}(x,\theta)\;=\;&
 \partial_{\gamma[K}\partial_{|\beta| L}\partial_{M|\alpha|}{\bf \Phi}_{IJ]}(x,\theta)\;=\;0,
\end{align}
which implies that the expansion of ${\bf \Phi}_{IJ}(x,\theta)$ in powers of  $\theta$ terminates at the
second order:
\begin{align}
\label{sol-Phi}
 {\bf \Phi}_{IJ}(x,\theta)\;=\;&\varphi_{IJ}(x)+\epsilon_{IJPQ}\theta^{\alpha P}\tilde\psi^Q_{\alpha}(x)
 +{1\over2}\epsilon_{IJPQ}\theta^{\alpha P}\theta^{\beta Q}\tilde f_{\alpha\beta}(x).
\end{align}
Eqns.~\eqref{const1-for-Psi},~\eqref{const2-for-Psi} determine ${\bf \Psi}_{\dot\alpha J}(x,\theta)$. The solution only 
exists when
\begin{align}
\label{conds-1}
 \partial_{[\alpha|\dot\alpha}\tilde\psi_{|\beta]}^Q\;=\;&0  \;\text{and} \;\; \partial_{[\alpha|\dot\alpha}\tilde f_{|\beta]\gamma}\;=\;0. 
\end{align}
which are Dirac and Maxwell equations. It is given by the following expression:
\begin{align}
\label{sol-Psi}
 {\bf \Psi}_{J\dot\alpha}(x,\theta)\;=\;&\psi_{J\dot\alpha}(x)+ \theta^{\alpha I}\partial_{\alpha\dot\alpha}\varphi_{JI}(x)+
 {1\over2}\epsilon_{JIPQ}\theta^{\alpha I}\theta^{\beta P}\partial_{(\alpha|\dot\alpha}\tilde\psi_{|\beta)}^Q(x)  \nonumber \\
 \;+\;& {1\over 3!}\epsilon_{JIPQ}\theta^{\alpha I}\theta^{\beta P}\theta^{\gamma Q}\partial_{(\alpha|\dot\alpha}\tilde f_{|\beta\gamma)}(x),
\end{align}
Eq.~\eqref{const2-in-F+} determines ${\bf F}^+_{\dot\alpha\dot\beta}(x,\theta)$, but the solution only exists when besides Eqs.~\eqref{conds-1} also:
\begin{align}
 \Box\varphi_{IJ}\;=\;0,
\end{align}
It is given by:
\begin{align}
\label{sol-F+}
 {\bf F}^+_{\dot\alpha\dot\beta}(x,\theta)\;=\; &f_{\dot\alpha\dot\beta}(x)+\theta^{J\rho}\partial_{\rho\dot\beta}\psi_{J\dot\alpha}(x)
  +{1\over2} \theta^{J\rho}\theta^{\alpha I}\partial_{(\rho|\dot\beta}\partial_{|\alpha)\dot\alpha}\varphi_{IJ}(x)  \nonumber \\
  \;+ \; &
 {1\over3!}\epsilon_{JIPQ}\theta^{J\rho}\theta^{\alpha I}\theta^{\beta P}\partial_{(\rho|\dot\beta|}\partial_{\alpha|\dot\alpha|}\tilde\psi_{\beta)}^Q(x) \nonumber \\
 \;+\; &
 {1\over4!}\epsilon_{JIPQ}\theta^{J\rho}\theta^{\alpha I}\theta^{\beta P}\theta^{\gamma Q}\partial_{(\rho|\dot\beta|}\partial_{\alpha|\dot\alpha|}\tilde f_{\beta\gamma)}(x).
\end{align}
To conclude, $\sigma$ is given by Eqn.~\eqref{def-of-sigma} where ${\bf F}^+_{\dot\alpha\dot\beta}(x,\theta),\; {\bf \Psi}_{\dot\beta J}(x,\theta)$ 
and ${\bf \Phi}_{IJ}(x,\theta)$ are given by Eqns.~\eqref{sol-F+},~\eqref{sol-Psi} and~\eqref{sol-Phi}, respectively.

\section{Coherent wavefunctions in free theory}\label{sec:CoherentWaveFunctions}
We will now discuss a special family of solutions of free equations, which form an orbit
of $PSL(4|4)$.

\vspace{10pt}
\noindent
For every point $Z\in {\bf T}$ we define a generalized function section of $\left(|\mbox{Ber}|\;{\cal S}_M\right)^{-1}$
(see Eq. (\ref{TypeOfSigma})), which we call $\delta_Z$. By definition $\delta_Z(X)$, for a fixed $2|2$-dimensional
$X\subset {\bf T}$,  is the 
delta-function of the condition that $Z\in X$. More precisely
(here $\Gamma(\mbox{\tt line bundle})$ denotes the space of sections): \marginpar{$\delta_Z(X)$}
\begin{align}
\delta_Z\;\in\; & \Gamma \left(|\mbox{Ber}|\;{\cal S}_M\right)^{-1} &
\nonumber \\  
& \mbox{\tt\small such that }\forall\; s \;\in\; \Gamma\left(|\mbox{Ber}|\;{\cal S}_M\right)\;:
\nonumber \\    
& \int_{Z\in \bf T}  f(Z) \langle s(X),\delta_Z(X)\rangle \;=\; \int_{Z\in X} s(X) f(Z),
\end{align}
Here the integration on the left hand side  uses the canonical measure on ${\bf T}$, and the 
integration measure on the right hand side is the integration along the fiber of ${\cal S}_M$
defined by the volume element $s$ (Section \ref{sec:Superdeterminant}).
In coordinates (\ref{CoordinatesInTwistorSpace}), (\ref{BasisOfL}):
\begin{align}
\delta_Z(X) \;=\; &
(u_1^I\zeta_I +  \tilde{\theta}^{\dot{\alpha}}_1\lambda_{\dot{\alpha}}) 
(u_2^J\zeta_J +  \tilde{\theta}^{\dot{\alpha}}_2\lambda_{\dot{\alpha}}) \;\times\; 
\nonumber\\    
& \times \delta(\mu^1 + x^{1\dot{\alpha}}\lambda_{\dot{\alpha}} + \theta^{1 I}\zeta_I) \,
\delta(\mu^2 + x^{2\dot{\alpha}}\lambda_{\dot{\alpha}} + \theta^{2 I}\zeta_I) 
\end{align}
So defined $\delta_Z(X)$, as a function of $X\in M$ (or rather a section of the line bundle 
$\left(|\mbox{Ber}|\,{\cal S}\right)^{-1}$ over $M$), encodes a solution of ${\cal N}=4$ super-Maxwell equations of motion
as described in Section \ref{sec:HarmonicSuperspace}. 

To determine the corresponding scalar, electromagnetic and spinor fields, we write 
$\delta_Z(X)$ as a sum of expressions with definite values of the $R$ charge, which corresponds to 
twice the power of $\zeta$:
\begin{align}
\tilde{\theta}^{\dot{\alpha}}_1\tilde{\theta}^{\dot{\beta}}_2
f_{\dot{\alpha}\dot{\beta}}\;:\;\; & 
\tilde{\theta}^{\dot{\alpha}}_1\tilde{\theta}^{\dot{\beta}}_2\lambda_{\dot{\alpha}}\lambda_{\dot{\beta}}
\delta^2(\mu + x\lambda) 
\\   
u_{[1}^J\tilde{\theta}_{2]}^{\dot{\alpha}} \psi_{\dot{\alpha} J}\;:\;\; &
\tilde{\theta}_{[1}^{\dot{\alpha}}\lambda_{\dot{\alpha}}u_{2]}^J\zeta_J
\delta^2(\mu + x\lambda) 
\\   
u_1^Iu_2^J\varphi_{IJ}\;:\;\; &
u_1^Iu_2^J\zeta_I\zeta_J
\delta^2(\mu + x\lambda) 
\label{PhiIJ}\\   
u_1^{[I}u_2^J\theta^{K]\alpha}\tilde{\psi}_{IJK\alpha}\;:\;\; &
u_1^{[I}u_2^J\theta^{K]\alpha}\zeta_I\zeta_J\zeta_K\partial_{\alpha}\delta^2(\mu + x\lambda)
\\   
u_1^{[I}u_2^J\theta^{K\alpha}\theta^{L]\beta}\tilde f_{\alpha\beta}\;:\;\;&
u_1^{[I}u_2^J\theta^{K\alpha}\theta^{L]\beta}\zeta_I\zeta_J\zeta_K\zeta_L\partial_{\alpha}\partial_{\beta}\delta^2(\mu + x\lambda)
\end{align}

\section{Beta deformation of the free Yang-Mills action}\label{sec:betadef}
We want to characterize  a  deforming operator $\int d^4 x \,U$ in Eq. (\ref{DeformationsOfSYM}) by its value on free solutions. 
In particular, in this paper we are interested in the case of beta-deformation,
which corresponds to deforming the action by an expression cubic in the elementary fields \cite{Milian:2016xuy}.  
Let us evaluate the corresponding $U$ on the formal sum of three coherent states: 
\begin{equation}\label{Perturbiner}
\left(\varepsilon_1\delta_{Z_{(1)}}(X)+ \varepsilon_2\delta_{Z_{(2)}}(X) + \varepsilon_3\delta_{Z_{(3)}}(X)\right)M
\end{equation} 
with bosonic nilpotent coefficients\footnote{``Nilpotent'' means satisfying
  $\varepsilon_1^2 = \varepsilon_2^2  = \varepsilon_3^2 = 0$.
  To the best of our knowledge, the idea to use nilpotent coefficient was first suggested
  in \cite{Rosly:1996vr}. The authors of \cite{Rosly:1996vr} constructed a solution
  of classical field equations, which they called ``perturbiner''. Our use of nilpotent
  coefficients is slightly different; we take a solution of the {\em free} field equations
  in the form of Eq. (\ref{Perturbiner}) and evaluate some $\int d^4x\;U$ on it. We need the
  nilpotence of coefficients to avoid considering the square of delta-function.}
 $\varepsilon_1$, $\varepsilon_2$, $\varepsilon_3$, where $M$ is some $N\times N$-matrix
(see Eq. (\ref{MaxwellTimesMatrix})).
As we explained in Section \ref{sec:Introduction}, linearized beta-deformations are parametrized by
$B \in ({\bf g}\wedge {\bf g})_0/{\bf g}$. We consider the case when $B$ is a decomposable tensor, {\it i.e.} is
of the form $B = \xi\wedge \xi$ \marginpar{$\xi\wedge\xi$}
where $\xi$ is an odd element of $\bf g$. The condition of zero internal
commutator (the subindex $0$ in $({\bf g}\wedge {\bf g})_0$) means that $\xi$ should be nilpotent.
We claim that the result of evaluation of $\int d^4x\; U$ on the free solution given by Eq. (\ref{Perturbiner})
is:
\begin{equation}\label{EvaluationEqualsDelta}
 \int d^4x\; U \;=\;  \varepsilon_1\varepsilon_2\varepsilon_3\,\delta^{\xi}_{Z_{(1)},Z_{(2)}}(Z_{(3)})\;\mbox{tr}M^3
\end{equation}
where $\delta^{\xi}_{Z_{(1)},Z_{(2)}}(Z_{(3)})$ is defined by Eq. (\ref{DefDelta}). In fact,
our Eq. (\ref{DefDelta}) works for any nilpotent $\xi\in \Pi T_{e} PSL(4|4,{\bf R})$; this includes
odd elements of $psl(4|4,{\bf R})$ as well as
linear combinations of even elements of $psl(4|4,{\bf R})$ with Grassmann odd coefficients\footnote{Since we work with coherent states, we are into supergeometry/supermanifolds; 
we have to use a ``pool'' of constant Grassmann odd parameters \cite{Shvarts:1985fe}.}.

\vspace{12pt}
\noindent
(At the same time, {\em on the string theory side} the integrated vertex operator of the
worldsheet sigma-model for this particular $B$ is \cite{Bedoya:2010qz}:
\begin{equation}\label{IntegratedVertex}
U_{\rm AdS} = \int j_a\xi^a \wedge j_b\xi^b
\end{equation}
Here $j_a$ is the Noether current on the worldsheet, a one-form.)

\subsection{Definition of $\delta^{\xi}_{Z_{(1)},Z_{(2)}}(Z)$}
Let us pick two points\footnote{Strictly speaking, it is not appropriate to think about ``points
  of a supermanifold''. It is better to say that, for an arbitrary supermanifold $S$, we
  pick two arbitary morphisms $Z_{(1)}:S\rightarrow {\bf PT}$ and $Z_{(2)}:S\rightarrow {\bf PT}$
  \cite{BernsteinLecture1}.
 For our purposes, it is enough to take $S={\bf R}^{0|K}$ for large enough $K$.
  Technically, we just allow all our twistors to depend on $K$ constant Grassmann odd parameters.
}  in ${\bf PT}$: $[Z_{(1)}]$ and $[Z_{(2)}]$. Let $\xi\in {\bf g}$ be an {\bf odd nilpotent} 
element of ${\bf g} = psl(4|4,{\bf R})$. We can represent $\xi$ as a $4|4\times 4|4$-supermatrix,
which we also denote $\xi$. Since $\xi$ is nilpotent as an element of $psl(4|4,{\bf R})$, the
square of this matrix is proportional to the unit matrix:
\begin{equation}
\xi^2 = c{\bf 1}
\end{equation}
where $c$ is some number. We assume that $\xi$ is nondegenerate, in the
following sense: either $c\neq 0$, or if $c=0$ then $\mbox{Ker}\,\xi=\mbox{Im}\,\xi$.
Let us consider a $2|2$-dimensional plane $L$ generated by $Z_{(1)}$, $Z_{(2)}$, $\xi Z_{(1)}$ and $\xi Z_{(2)}$:
\begin{equation}\label{LGenerated}
L \;=\; {\bf R}Z_{(1)} + {\bf R} Z_{(2)} + {\bf R}\xi Z_{(1)} + {\bf R}\xi Z_{(2)}
\end{equation}
Let us define $\delta^{\xi}_{Z_{(1)},Z_{(2)}}(Z)$, essentially a delta-function of $Z$, in the following way.
For any test function $f\in C^{\infty}({\bf T})$:\marginpar{$\delta_{Z_{(1)},Z_{(2)}}^{\xi}$}
\begin{align}
  & \int_{Z\in{\bf R}^{4|4}} \delta^{\xi}_{Z_{(1)},Z_{(2)}}(Z)\, f(Z) \;=\; 
    \nonumber\\   
  \;=\;
  & \int_{{\bf R}^2} da_1\wedge da_2\;{\partial\over\partial\psi^1}{\partial\over\partial\psi^2}\;
    f(a_1 Z_{(1)} + a_2 Z_{(2)} + \psi^1\xi Z_{(1)} + \psi^2 \xi Z_{(2)})
\label{DefDelta}
\end{align}
The integration on the left hand side uses the canonical measure on ${\bf R}^{4|4}$
(the ``$S$'' in ``$PSL(4|4)$''). The integral on the right hand side uses the integral form
on $L$ canonically defined  by $\xi$ as described in Appendix \ref{sec:Measure}.

Eq. (\ref{DefDelta}) implies that
$\delta^{\xi}_{Z_{(1)},Z_{(2)}}(Z)$ is a linear function of $\xi\otimes\xi$. It has weight zero in $Z$
and also in both $Z_{(1)}$ and $Z_{(2)}$; in other words, for $\kappa\in {\bf R}$:
\begin{equation}
\delta^{\xi}_{Z_{(1)},Z_{(2)}}(\kappa Z) = \delta^{\xi}_{\kappa Z_{(1)},Z_{(2)}}(Z) = \delta^{\xi}_{Z_{(1)},\kappa Z_{(2)}}(Z) =\delta^{\xi}_{Z_{(1)},Z_{(2)}}(Z)
\end{equation}
Equivalently, we can define $\delta^{\xi}_{Z_{(1)},Z_{(2)}}(Z)$ as follows:
\begin{align}
  & \delta^{\xi}_{Z_{(1)},Z_{(2)}}(Z) \;=\;
    \label{IntegratedDeltaFunction}
  \\   
  \;=\;
  & \int_{{\bf R}^2} da_1\wedge da_2\;{\partial\over\partial\psi^1}{\partial\over\partial\psi^2}\;
    \delta^{(4|4)}(Z - a_1 Z_{(1)} - a_2 Z_{(2)} - \psi^1\xi Z_{(1)} - \psi^2 \xi Z_{(2)})
    \nonumber
\end{align}

\subsection{Orientation of $L_{\rm rd}$ and discontinuity of $\delta^{\xi}_{Z_{(1)},Z_{(2)}}$}
In order to integrate an integral form:
\begin{equation}
da_1\wedge da_2\;{\partial\over\partial\psi^1}{\partial\over\partial\psi^2}\;
f(a_1 Z_{(1)} + a_2 Z_{(2)} + \psi^1\xi Z_{(1)} + \psi^2 \xi Z_{(2)})
\end{equation}
over a supermanifold $L$ (Eq. (\ref{LGenerated})),
we need an {\em orientation} of its body $L_{\rm rd} = {\bf R}^2$. This ${\bf R}^2$ is generated by $Z_{(1)}$ and $Z_{(2)}$.
Therefore, we need to know which basis has positive orientation:
$\{Z_{(1)},Z_{(2)}\}$ or $\{Z_{(2)},Z_{(1)}\}$? A  comparison with explicit calculation in Section \ref{sec:PhiPsiPsi} shows
that the orientation is determined by the sign of $\epsilon^{\dot{\alpha}\dot{\beta}}\lambda_{(1)\dot{\alpha}}\lambda_{(2)\dot{\beta}}$. This sounds like a
contradiction, because  $\epsilon^{\dot{\alpha}\dot{\beta}}\lambda_{(1)\dot{\alpha}}\lambda_{(2)\dot{\beta}}$ is not conformally invariant. However, the {\em sign}
of  $\epsilon^{\dot{\alpha}\dot{\beta}}\lambda_{(1)\dot{\alpha}}\lambda_{(2)\dot{\beta}}$ is invariant under {\em infinitesimal} conformal transformations.
Explicit calculation in Section \ref{sec:PhiPsiPsi} shows that $\delta^{\xi}_{Z_{(1)},Z_{(2)}}(Z)$, as a function of $Z_{(1)}$ and $Z_{(2)}$,
has a discontinuity when $\epsilon^{\dot{\alpha}\dot{\beta}}\lambda_{(1)\dot{\alpha}}\lambda_{(2)\dot{\beta}}=0$. The limit at the discontinuity is proportional
to $\delta^{(2)}(\lambda - a_1\lambda_{(1)} - a_2\lambda_{(2)})$, but the sign in front of $\delta^{(2)}(\lambda - a_1\lambda_{(1)} - a_2\lambda_{(2)})$
depends on whether we are approaching from  $\epsilon^{\dot{\alpha}\dot{\beta}}\lambda_{(1)\dot{\alpha}}\lambda_{(2)\dot{\beta}}>0$
or from  $\epsilon^{\dot{\alpha}\dot{\beta}}\lambda_{(1)\dot{\alpha}}\lambda_{(2)\dot{\beta}}<0$.

This can be explained as follows. We interpret the four-dimensional space-time as the real
Grassmannian: ${\bf R}^{2,2}\cup \infty=\mbox{Gr}(2,4)$ (Section \ref{sec:HarmonicSuperspace}). For the Yang-Mills action to be
conformally invariant, the scalar field should transform as a section of $(\mbox{Ber}S)^{-1}$,
and spinors as sections of $S\otimes (\mbox{Ber}\, S)^{-1}$ (chiral) and $S^{\perp}\otimes (\mbox{Ber}\,S)^{-1}$ (antichiral),
where $S$ is the
tautological vector bundle of $\mbox{Gr}(2,4)$ (like ${\cal S}_M$ of Section \ref{sec:TwistorsAndGrassmannian} but without ``super-'').
However, the $\delta_Z$ of Section \ref{sec:CoherentWaveFunctions} gives sections of $(\ldots)\otimes|\mbox{Ber}|^{-1}$
instead of $(\ldots)\otimes\mbox{Ber}^{-1}$
(Section \ref{sec:Superdeterminant}). We would be able to cast sections of  $|\mbox{Ber}|^{-1}$ as sections of $\mbox{Ber}^{-1}$,
if we provided an orientation of $S$. But this is only possible locally. In fact, we can
provide orientation of the fiber of $S$ over all $x$ except at infinity.
This is acceptable for our computation, because the integral $\int d^4x \;U$ in Eq. (\ref{DeformationsOfSYM}) is
supported on a single point  $x\in {\bf R}^{2,2}$ determined by the intersection of the planes
$\mu_{(1)}-\lambda_{(1)} x = 0$ and $\mu_{(2)} -\lambda_{(2)}x=0$. But when we consider the case when
$\lambda_{(1)}\rightarrow \lambda_{(2)}$, the intersection point $x$ goes to $\infty$, where the orientation is undefined,
essentially because $||x||^2 = \mbox{det}\,x$ can be positive or negative depending on how $x\to\infty$.
This leads to the discontinuity of $\delta^{\xi}_{Z_{(1)},Z_{(2)}}(Z)$.

\subsection{Symmetry of  $\delta^{\xi}_{Z_{(1)},Z_{(2)}}(Z)$}\label{sec:SymmetryOfDelta}
The defining Eq. (\ref{DefDelta}) implies that $\delta^{\xi}_{Z_{(1)},Z_{(2)}}(Z)$ is symmetric under the exchange
$Z_{(1)}\leftrightarrow Z_{(2)}$. In fact it is symmetric under arbitrary permutations of $Z,Z_{(1)},Z_{(2)}$. This
can be proven as follows:
\begin{align}
  & \delta^{(4|4)}(Z - a_1 Z_{(1)} - a_2 Z_{(2)} - \psi_1\xi Z_{(1)} - \psi_2 \xi Z_{(2)})
    \;=\;
  \\
  \;=\;
  &\mbox{sign}(a_1) \delta^{(4|4)}\left(
    {1\over a_1 + \psi_1\xi}Z - Z_{(1)} - {a_2\over a_1 + \psi_1\xi}Z_{(2)} -
    {1\over a_1 + \psi_1\xi}\psi_2\xi Z_{(2)}
    \right)\;=
  \nonumber\\
  \;=\;
  &\mbox{sign}(a_1)\times
  \\
  & \times \delta^{(4|4)}\Big(
    -Z_{(1)} + a_1^{-1} Z - (a_1^{-1}a_2 - a_1^{-2}\psi_1\psi_2c) Z_{(2)} - a_1^{-2}\psi_1\xi Z \; -
  \\
  & \phantom{\times\delta^{(4|4)}\Big(\;\;}
    -\;(a_1^{-1}\psi_2 - a_1^{-2}a_2\psi_1)\xi Z_{(2)}
    \Big)
    \nonumber
\end{align}
The change of variables:
\begin{align}
  \tilde{a}_1 = & \; a_1^{-1}
  \\
  \tilde{a}_2 = & - a_1^{-1}a_2 + a_1^{-2}\psi_1\psi_2c
  \\
  \tilde{\psi}_1 = & - a_1^{-2}\psi_1
  \\
  \tilde{\psi}_2 = & \; a_1^{-2}a_2\psi_1 - a_1^{-1}\psi_2
\end{align}
in the integral $\int_{{\bf R}^2} da_1\wedge da_2\;{\partial\over\partial\psi^1}{\partial\over\partial\psi^2}\;$is equivalent to the exchange\footnote{The sign
  factor $\mbox{sign}(a_1)$ is compensated by the possible difference of sign between $\epsilon^{\dot{\alpha}\dot{\beta}}\lambda_{(1)\dot{\alpha}}\lambda_{(2)\dot{\beta}}$ and $\epsilon^{\dot{\alpha}\dot{\beta}}\lambda_{\dot{\alpha}}\lambda_{(2)\dot{\beta}}$. This
sign determines the orientation of ${\bf R}^2$ which is needed to integrate $da_1\wedge da_2$.} $Z\leftrightarrow Z_1$.

\subsection{The case when $\xi\wedge \xi$ corresponds to zero deformation}\label{sec:ZeroDeformations}
Suppose that $\xi\wedge\xi$ belongs to the denominator of Eq. (\ref{Representation}):
\begin{equation}
\xi\wedge\xi = \Delta(\zeta) = \sum C^{ab}t_a\wedge [t_b,\zeta]
\end{equation}
for some $\zeta\in {\bf psl(4|4)}$, and $C^{ab}$ the inverse of the Killing metric. We know from \cite{Bedoya:2010qz} that
such $\xi\wedge\xi$ should correspond to zero deformations. We will now prove that for such $\xi$ Eq. (\ref{DefDelta}) 
indeed evaluates to zero.

Let us enumerate the basis vectors of ${\bf T}$: $\{ e_1, e_2, e_3, e_4 , f_1, f_2, f_3, f_4\}$
where $e$ are even and $f$ are odd. Without loss of generality, let us assume that
$Z_{(1)} = e_1$ and $Z_{(2)} = e_2$. (In other words, we choose a basis so that the first
two elements are $Z_{(1)}$ and $Z_{(2)}$.) 

Let us first consider the case when $\zeta$ is even:
\begin{equation}
\zeta = e_a\otimes e^{\vee b} - {1\over 4} \delta_a^b {\bf 1}
\end{equation}
where $e^{\vee},f^{\vee}$ are elements of the dual basis and ${\bf 1}  = \sum_c e_c\otimes e^{\vee c}$. In this case:
\begin{align}
\Delta(\zeta) \;=\; & 
\sum_{c=1}^4\left(e_a\otimes e^{\vee c} - {1\over 4}\delta_a^c {\bf 1}\right)\wedge 
\left(e_c\otimes e^{\vee b} - {1\over 4}\delta_c^b {\bf 1}\right) + 
\mbox{odd}\wedge\mbox{odd}
\end{align}
The ${\rm odd}\wedge{\rm odd}$ terms do not contribute to the computation. Also, the subtractions do not
contribute to the computation, because the integral of Eq. (\ref{IntegratedDeltaFunction}):
\begin{equation}
   \int_{{\bf R}^2} da^1\wedge da^2 {\partial\over\partial\psi^1}{\partial\over\partial\psi^2}
   f(a^1Z_{(1)} + a^2 Z_{(2)} + \psi^1\xi_1 Z_{(1)} + \psi^2 \xi_2 Z_{(2)})
\end{equation}
is zero when either $\xi_1$ or $\xi_2$ is proportional to $\bf 1$ (for example, when $\xi_1 = \alpha {\bf 1}$,
the dependence on $\psi^1$ can be removed by shifting $a^1 \mapsto a^1 - \psi^1\alpha$).
Let us pick a pair of fermionic constants $\epsilon$ and $\eta$ from the pool\cite{Shvarts:1985fe}; we are evaluating:
\begin{align}
\int_{{\bf R}^2} da_1\wedge da_2{\partial\over\partial\psi^1}{\partial\over\partial\psi^2} \sum_{c=1}^4 f\Big( & 
a^1e_1 + a^2 e_2 \;+\; 
\nonumber\\   
& 
+ \psi^1 (\epsilon \delta^c_1 e_a + \eta \delta^b_1 e_c)  
+ \psi^2 (\epsilon \delta^c_2 e_a + \eta \delta^b_2 e_c)
\Big)
\end{align}
This integral is equal to zero. In fact, each term in $\sum_{c=1}^4$ vanishes separately. 
Indeed, when $c$ is $3$ or $4$ it is proportional to $\eta^2=0$. 
If $c$ is $1$ or $2$, we can shift the integration variable $a^1$ or $a^2$ to eliminate the 
dependence on $\eta$; then the integral becomes proportional to $\epsilon^2=0$.

Now consider the case when $\zeta$ is odd:
\begin{align}
\zeta \;=\; & f_a\otimes e^{\vee b}
\\    
\Delta(\zeta) \;=\; &
(f_a\otimes e^{\vee c})\wedge \left(e_c \otimes e^{\vee b} - {1\over 4}\delta_c^b{\bf 1}\right) + 
\left(f_a\otimes f^{\vee c} - {1\over 4}\delta^c_a{\bf 1}\right)\wedge (f_c \otimes e^{\vee b})
\end{align}
Again, the second term does not contribute to the computation. The contribution of the first 
term is:
\begin{align}
  \int_{{\bf R}^2} da_1\wedge da_2{\partial\over\partial\psi^1}{\partial\over\partial\psi^2}
  \sum_c f\Big(
  & a^1e_1 + a^2 e_2 \;+\;
  \\   
  & 
    + \psi^1 (\delta^c_1 f_a + \eta \delta^b_1 e_c)  
    + \psi^2 (\delta^c_2 f_a + \eta \delta^b_2 e_c)
    \Big)
    \nonumber
\end{align}
The terms with $c$ equal $3$ or $4$ are zero, being proportional to $\eta^2=0$. When $c$ is $1$ or $2$,
we can eliminate the dependence on $\eta$ by shifting $a$, and then we are left with, for example, 
when $c=1$:
\begin{equation}
\int_{{\bf R}^2} da_1\wedge da_2{\partial\over\partial\psi^1}{\partial\over\partial\psi^2}
\sum_c f\Big( 
a^1e_1 + a^2 e_2 + \psi^1  f_a 
\Big)
\end{equation}
This integral is zero, because the integrand does not depend on $\psi^2$.

\subsection{Deformation of the form $\int \phi\psi\psi$}\label{sec:PhiPsiPsi}
To prove our claim that the evaluation of the deforming operator on the free solution given
by Eq. (\ref{Perturbiner}) is indeed $\delta^{\xi}_{Z_{(1)},Z_{(2)}}(Z)$, we will consider a particular case when the test
function $f(Z)$ of Eq. (\ref{DefDelta}) is: 
\begin{equation}
f(Z) \;=\; A_{IK}(\lambda,\mu) \epsilon^{IKPQ}\zeta_P\zeta_Q
\end{equation}
Eq. (\ref{EvaluationEqualsDelta}) implies that evaluation of $\int_{Z\in{\bf R}^{4|4}} \delta^{\xi}_{Z_{(1)},Z_{(2)}}(Z)\, f(Z)$ with such $f(Z)$
corresponds to the substitution instead of the term proportional to  $\varepsilon_3$ in Eq. (\ref{Perturbiner}) 
of the following solution of the linearized equations of motion ({\it cp.} Eq. (\ref{PhiIJ})):
\begin{align}
& \varphi_{IK}(x)\;=\;\int d^2\lambda d^2\mu \; A_{IK}(\lambda,\mu)\delta^{2}(\mu + x\lambda)
\label{PhiVsA}\\    
& \psi = \tilde{\psi} = F^+ = F^- \;=\; 0
\end{align}
Let us choose both $\xi_1$ and $\xi_2$ as ``rotations of $S^5$'', {\it i.e.}:
\begin{equation}
(\xi_1\wedge\xi_2)^I_K{}^J_L = B^{IJ}_{KL}
\end{equation}
with other components of $\xi_1\wedge\xi_2$ all zero.

\vspace{10pt}
\noindent
Eq. (\ref{DefDelta}) implies that $\int_{Z\in{\bf R}^{4|4}} \delta^{\xi}_{Z_{(1)},Z_{(2)}}(Z)\, f(Z)$ equals to:
\begin{equation}\label{AsIntegrationOverPlane}
 \int_{{\bf R}^2} A_{PQ}(a^{(1)}\lambda_{(1)} + a^{(2)}\lambda_{(2)}, a^{(1)}\mu_{(1)} + a^{(2)}\mu_{(2)}) 
\epsilon^{PQRS} B^{IJ}_{RS} \zeta_{(1)I} \zeta_{(2)J} da^{(1)}\wedge da^{(2)}
\end{equation}
The same result is obtained by evaluating the deformation term in the Lagrangian \cite{Milian:2016xuy},
using the formulas of Appendix \ref{sec:Integrals}:
\begin{align}
& \int d^4 x \epsilon^{IJKL}\varphi_{IJ}(x) B_{KL}^{PQ}
\psi_{(1)P\dot{\alpha}}(x) \psi_{(2)Q\dot{\beta}}(x)\epsilon^{\dot{\alpha}\dot{\beta}}
\label{LagrangianForBeta}\\    
\mbox{\tt\small where } & 
\psi_{(A)\dot{\alpha}K}(x) = \lambda_{(A)\dot{\alpha}}\zeta_{(A)K} \delta^{2}(\mu_{(A)}+x\lambda_{(A)})
\end{align}
This establishes the contact with the description of the supermultiplet on the field theory side 
obtained in \cite{Milian:2016xuy}; matching of other states follows from applying supersymmetry\footnote{Vanishing
of deformations corresponding to the denominator in Eq. (\ref{Representation}) was not proven
in \cite{Milian:2016xuy}. It follows from Section \ref{sec:ZeroDeformations} that such
deformations would vanish when evaluated on (\ref{Perturbiner}). This should imply that
the deformation is acually zero, although we do not have a rigorous proof.}.

\paragraph     {Fourier transform into the usual momentum space}
We will use now the notations of \cite{Witten:2003nn}. Following the prescription in \cite{Witten:2003nn}, let us substitute
in Eq. (\ref{PhiVsA}): 
\begin{equation}
   A_{IK}(\lambda,\mu) = \delta^2(\lambda - \lambda_{(0)})\exp\left(i\mu \tilde\lambda_{(0)}\right)
   \zeta_{(0)I}\zeta_{(0)K}
\end{equation}
and then multiply by the Fourier transform factor $\exp\left(-i\mu_{(1)}\tilde{\lambda}_{(1)}-i\mu_{(2)}\tilde{\lambda}_{(2)}\right)$
and integrate over $\mu_{(1)}$ and $\mu_{(2)}$. Then Eq. (\ref{AsIntegrationOverPlane}) gives:
\begin{align}
  \int da^1\wedge da^2\;
  &
   \delta^2\left(
      \lambda_{(0)} - a^{(1)}\lambda_{(1)} - a^{(2)}\lambda_{(2)}
    \right)\times
  \\
  \times
  &
    \delta^2\left(
    \tilde{\lambda}_{(1)} - a^{(1)}\tilde{\lambda}_{(0)}
    \right)\delta^2\left(
    \tilde{\lambda}_{(2)} - a^{(2)}\tilde{\lambda}_{(0)}
    \right)\;\zeta_{(0)P}\zeta_{(0)Q}\epsilon^{PQRS} B^{IJ}_{RS} \zeta_{(1)I} \zeta_{(2)J}
    \nonumber
\end{align}
Using Appendix \ref{sec:Integrals}, this is equal to:
\begin{align}
  & {1\over \langle\lambda_{(1)},\lambda_{(2)}\rangle }\;
    \delta^2\left(\tilde{\lambda}_{(1)} 
    - {\langle\lambda_{(2)},\lambda_{(0)}\rangle\over\langle\lambda_{(2)},\lambda_{(1)}\rangle}
    \tilde{\lambda}_{(0)}
    \right)
    \delta^2\left(\tilde{\lambda}_{(2)} 
    - {\langle\lambda_{(1)},\lambda_{(0)}\rangle\over\langle\lambda_{(1)},\lambda_{(2)}\rangle}
    \tilde{\lambda}_{(0)}
    \right)\times
  \nonumber\\   
  & \times \zeta_{(0)P}\zeta_{(0)Q}\epsilon^{PQRS} B^{IJ}_{RS} \zeta_{(1)I} \zeta_{(2)J}
    \label{AnswerInStandardNotations}
\end{align}
Remember \cite{Witten:2003nn} that the momenta of the scattering particles are $p_{(i)\alpha\dot{\alpha}} = \tilde{\lambda}_{(i)\alpha}\lambda_{(i)\dot{\alpha}}$.
Eq. (\ref{AnswerInStandardNotations}) can be interpreted as a deformation of the three-point scattering amplitude.
It is a generalized function with support on $\tilde{\lambda}_{(1)}$, $\tilde{\lambda}_{(2)}$ and $\tilde{\lambda}_{(0)}$ being all collinear to each
other.


\section{Other finite-dimensional representations}\label{sec:Generalization}
It is likely that the expressions $\delta^{\xi}_{Z_1,Z_2}(Z)$ will serve as building block for
evaluation of deformations corresponding to other finite-dimensional representations.
 It was conjectured in \cite{Mikhailov:2011af} that deformations of the form:
 \begin{align}
    & \int d^4 x \; \rho_0(x)\; \mbox{tr}(\Phi_1 + i\Phi_2)^4(x)\;,\;
      \int d^4 x \; \rho_1(x)\; \mbox{tr}(\Phi_1 + i\Phi_2)^5(x)\;,\;
   \nonumber\\
   & \int d^4x \;\rho_2(x)\;\mbox{tr}(\Phi_1 + i\Phi_2)^6(x)\;,\hspace{20pt}\ldots
     \label{SeriesOfOperators}
\end{align}
where $\rho_n$ are some special polynomial functions described in \cite{Mikhailov:2011af}
generate finite-dimensional representations of $\bf g$ corresponding to supersymmetric Young diagramms:
\begin{equation}
\begin{picture}(60,60)(0,2)
\multiput(0,20)(10,0){2}{
  \put(0,10){\framebox(10,10){}}
  \put(0,0){\framebox(10,10){}}
} 
\multiput(20,0)(10,0){2}{
  \put(0,10){\framebox(10,10){}}
  \put(0,0){\framebox(10,10){}}
} 
\end{picture}
,
\hspace{10pt}
\begin{picture}(60,60)(0,2)
\put(0,20){
  \put(0,20){\framebox(10,10){}}
  \put(0,10){\framebox(10,10){}}
  \put(0,0){\framebox(10,10){}}
} 
\put(10,20){
  \put(0,20){\framebox(10,10){}}
  \put(0,10){\framebox(10,10){}}
  \put(0,0){\framebox(10,10){}}
} 
\multiput(20,0)(10,0){3}{
  \put(0,10){\framebox(10,10){}}
  \put(0,0){\framebox(10,10){}}
} 
\end{picture}
,
\hspace{10pt}
\begin{picture}(60,60)(0,2)
\put(0,20){
  \put(0,30){\framebox(10,10){}}
  \put(0,20){\framebox(10,10){}}
  \put(0,10){\framebox(10,10){}}
  \put(0,0){\framebox(10,10){}}
} 
\put(10,20){
  \put(0,30){\framebox(10,10){}}
  \put(0,20){\framebox(10,10){}}
  \put(0,10){\framebox(10,10){}}
  \put(0,0){\framebox(10,10){}}
} 
\multiput(20,0)(10,0){4}{
  \put(0,10){\framebox(10,10){}}
  \put(0,0){\framebox(10,10){}}
}
\hspace{80pt}
,
\hspace{10pt}
\ldots
\end{picture}
\end{equation}
and their transposed.
An element of such a representation is a super-traceless tensor $B$ with $2(n+2)$ lower indices
and $2(n+2)$ upper indices. The symmetry type of the lower indices is determined by the lower
portion of the Young diagramms, and of the upper indices by the upper portion. For example,
the Young diagramm \includegraphics[scale=0.5]{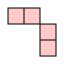} (``case $n=-1$'') corresponds to the
tensors of the form $B^{ab}_{cd}$ symmetric in $ab$ and antysymmetric in $cd$, while
\includegraphics[scale=0.5]{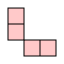} corresponds to the tensors of the form $B^{ab}_{cd}$
antisymmetric in $ab$ and symmetric in $cd$. When all indices $a, b, c, d$ are fermionic,
\includegraphics[scale=0.5]{beta.png}
gives $B^{IJ}_{KL} = B^{JI}_{KL} = - B^{IJ}_{LK}$ (see Eq. (\ref{LagrangianForBeta})).

Consider the tensor $B = \xi^{\otimes 2(n+2)}$ where
$\xi$ is an odd nilpotent element of $\bf g$. Let us apply the Young symmetrizer
corresponding to:
\begin{center}
   \begin{minipage}[c]{0.91\linewidth}
      \begin{center}
         \includegraphics[scale=0.25]{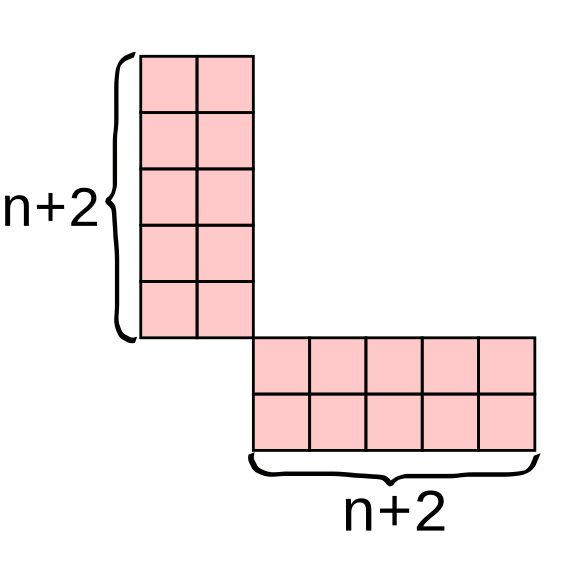}
      \end{center}
   \end{minipage}
   \begin{minipage}[c]{0.07\linewidth}
      {\refstepcounter{equation}\label{Young-projector}} 
      (\arabic{equation})
   \end{minipage}
\end{center}
We conjecture that the evaluation of the deformation on
the coherent state (as in Eq. (\ref{MaxwellTimesMatrix})):
\begin{equation}\label{BigPerturbiner}
   \left(
      \varepsilon_1\delta_{Z_1}(X)+ \varepsilon_2\delta_{Z_2}(X) +\ldots +
      \varepsilon_{n+4}\delta_{Z_{n+4}}(X)
   \right)M
\end{equation} 
gives $\delta^{\xi}_{Z_1,\ldots,Z_{n+4}}\mbox{tr}M^{n+4}$ where:
\begin{equation}\label{Generalization}
\delta^{\xi}_{Z_1,\ldots,Z_{n+4}}\;=\;\delta_{Z_1,Z_2}^{\xi}(Z_3)\delta_{Z_2,Z_3}^{\xi}(Z_4)\cdots \delta^{\xi}_{Z_{n+2},Z_{n+3}}(Z_{n+4})
\end{equation}
We leave the verification of this statement for future work.
The symmetry of $\delta^{\xi}_{Z_1,\ldots,Z_n}$ under permutations of $Z_i$ can be proven as in  Section \ref{sec:SymmetryOfDelta}.

Let us prove that Eq. (\ref{Generalization}) depends on $\xi\wedge\cdots\wedge\xi$ only through
its Young projector. To start, notice that each $\delta_{Z_n,Z_{n+1}}^{\xi}(Z_{n+2})$ depends on
$\xi\wedge\xi$ via the projector \includegraphics[scale=0.5]{beta-1.png} $\xi\otimes\xi$.
(In other words, the lower through $\xi_{[c}^{(a}\xi_{d]}^{b)}$.)
This follows from the fact that $\delta_{Z_n,Z_{n+1}}^{\xi}(Z_{n+2})=0$ when $\xi$ is degenerate,
$\mbox{dim im}\;\xi = 1$ (this follows from the definition Eq. (\ref{DefDelta})).
This implies that $\delta^{\xi}_{Z_1,\ldots,Z_{n+4}}$ depends on $\xi\wedge\cdots\wedge\xi$
only through:
\begin{equation}
   \xi_{[c_1}^{(a_1}\xi_{d_1]}^{b_1)}\;\xi_{[c_2}^{(a_2}\xi_{d_2]}^{b_2)}
   \cdots
   \xi_{[c_{n+2}}^{(a_{n+2}}\xi_{d_{n+2}]}^{b_{n+2})}
\end{equation}
But this is not all, there are more projectors. Let us consider the special case $n=0$.
Let us write $\delta^{\xi}_{Z_1,Z_2,Z_3,Z_{4}}$ as $\delta^{\xi}_{Z_2,Z_3}(Z_1)\delta^{\xi}_{Z_2,Z_3}(Z_4)$.
The indices $c_1$ and $c_2$ both contract with $Z_2$, therefore they enter symmetrized.
The indices $d_1$ and $d_2$ also enter symmetrized, because they both contract with $Z_3$.
Therefore, the $\xi\wedge\xi\wedge\xi\wedge\xi$ enters only through the projector
to \includegraphics[scale=0.5]{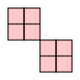}. In general, $\xi^{\wedge 2(n+2)}$ enters only through the
projector of Eq. (\ref{Young-projector}).

The second diagramm of Eq. (\ref{two-diagramms}) can be probed by replaying our construction with
$\bf T$ replaced with $\Pi \bf T$ (the twistor space with flipped statistics).

\section*{Acknowledgments}
The authors are partially supported by the 
FAPESP grant 2014/18634-9 ``Dualidade Gravitac$\!\!,\tilde{\rm a}$o/Teoria de Gauge''.
The work of A.M. was also supported in part by the 
RFBR grant 15-01-99504 ``String theory and integrable systems''. 
The work of S.P.M was supported by the CNPq grant 154704/2014-8. 

\appendix

\section{Measure defined by an odd linear operator}\label{sec:Measure}
A nondegenerate odd nilpotent linear operator $\xi$ on an $n|n$-dimensional linear space defines
an integration measure, in the following way.
Given {\em any} $n$-tuple of even vectors $Z_1,\ldots,Z_n$ we define the
integral of an arbitrary function $f$ as follows:
\begin{align}\label{IntF}
  \int f = \int_{{\bf R}^{n|n}}
  & da^1\cdots da^n
    {\partial\over\partial\psi^1}\cdots{\partial\over\partial\psi^n}
    f\left( a^iZ_i+ \psi^i\xi Z_i \right)
\end{align}
In fact, it is not strictly necessary that $\xi$ is nilpotent; it is enough that the square
of $\xi$ be proportional to a unit matrix:
\begin{equation}
\xi^2 = c{\bf 1}
\end{equation}
where $c$ is some number.
All we have to prove that the definition given by Eq. (\ref{IntF}) is independent of the choice
of $Z_1,\ldots,Z_n$. Suppose that we choose a different $n$-tuple:
\begin{equation}
\widetilde{Z}_i = Z_i + \epsilon_i^j \xi Z_j
\end{equation}
We replace  $Z$ with $\widetilde{Z}$ in Eq. (\ref{IntF}); the integrand becomes:
\begin{align}
  f\left( a^iZ_i+ \psi^i\xi Z_i \right)\;=\;
  & f\left(
     a^iZ_i + a^i\epsilon_i^j\xi Z_j + \psi^i\xi Z_i + \psi^i \epsilon^j_i c Z_j
     \right) \;=\;
  \label{NewIntegrand}\\
  \;=\;
   & f\left(
     a^iZ_i + (\psi^i + a^j\epsilon_j^i)\xi Z_i + (\psi^i + a^j\epsilon_j^i) \epsilon^k_i c Z_k
     - a^j\epsilon_j^k\epsilon_k^icZ_i
     \right)
     \nonumber
\end{align}
The difference between Eq. (\ref{NewIntegrand}) and the integrand of Eq. (\ref{IntF}) can
be undone by the change of variables $\widetilde{\psi}^i = \psi^i + a^j\epsilon_j^i$,
$\widetilde{a}^i = a^i - a^j\epsilon_j^k\epsilon_k^ic$. (Notice that the determinant of the
change from $a$ to $\widetilde{a}$ is $1$, since $\epsilon^i_j$ are fermionic.)

\section{Some integrals}\label{sec:Integrals}
We denote:
\begin{equation}
   \langle\lambda_1,\lambda_2\rangle = \epsilon^{\dot{\alpha}\dot{\beta}}\lambda_{1\dot{\alpha}}\lambda_{2\dot{\beta}}\quad,\quad
   \langle\mu_1,\mu_2\rangle = \epsilon_{\alpha\beta}\mu^{1\alpha}\mu^{2\beta}
\end{equation}
To integrate a 2-form $da^1\wedge da^2$, we need an orientation of the $(a_1,a_2)$-plane.
We orient it as $(a_1,a_2)$ if $\langle\lambda_1,\lambda_2\rangle >0$ and as $(a_2,a_1)$ otherwize. Then we get:
\begin{equation}
   \langle\lambda_1,\lambda_2\rangle \int da^1\wedge da^2\; \delta^2(\lambda - a^1\lambda_1 - a^2\lambda_2) = 1
\end{equation}
Therefore:
\begin{align}
  & \int d^4 x \; \delta^2 (\mu+ x\lambda) \;\langle\lambda_1,\lambda_2\rangle \;
    \delta^2(\mu_1 + x\lambda_1) \delta^2(\mu_2 + x\lambda_2)\;=
  \nonumber\\   
  =\;
  & \langle\lambda_1,\lambda_2\rangle^2
    \int da^1\wedge da^2\; \delta^2(\lambda - a^1\lambda_1 - a^2\lambda_2)\times
  \nonumber\\
  & \times
    \int d^4x\;
    \delta^2 (\mu+ x\lambda) \delta^2(\mu_1 + x\lambda_1) \delta^2(\mu_2 + x\lambda_2)\;=
  \\   
  =\;
  & \int da^1\wedge da^2\;
    \delta^2(\lambda - a^1\lambda_1 - a^2\lambda_2)\delta^{2}(\mu - a^1\mu_1 - a^2\mu_2)\;=
  \\   
  =\;
  & \langle\lambda_1,\lambda_2\rangle\;
    \delta^2 \Big(
    \langle \lambda_1,\lambda_2 \rangle \mu
    - \langle\lambda,\lambda_2\rangle\mu_1
    + \langle\lambda,\lambda_1\rangle\mu_2
    \Big)=
  \\   
  =\;
  & \langle\lambda_1,\lambda_2\rangle\;\left|\langle\mu_1,\mu_2\rangle\right|\;
    \delta\Big(
    \langle \lambda_1,\lambda_2 \rangle \langle\mu,\mu_1\rangle
    - \langle\mu_1,\mu_2\rangle \langle \lambda,\lambda_1 \rangle
    \Big)\times
  \nonumber\\
  & \times\delta\Big(
  \langle \lambda_1,\lambda_2 \rangle \langle\mu,\mu_2\rangle
  - \langle\mu_1,\mu_2\rangle \langle \lambda,\lambda_2 \rangle
  \Big)
\end{align}


\def\cprime{$'$} \def\cprime{$'$}
\providecommand{\href}[2]{#2}\begingroup\raggedright\endgroup

\end{document}